\documentstyle[preprint,tighten,prc,aps,epsfig]{revtex}

\newcommand{\be}{\begin{equation}}
\newcommand{\ee}{\end{equation}}
\newcommand{\bfeps}{\mbox {\boldmath $\epsilon$}}
\newcommand{\bfp}{\mbox {\boldmath $p$}}
\newcommand{\bfk}{\mbox {\boldmath $k$}}
\newcommand{\bfq}{\mbox {\boldmath $q$}}

\newcommand{\bfg}{\mbox {\boldmath $G$}}
\newcommand{\bfn}{\mbox {\boldmath $n$}}
\newcommand{\bfr}{\mbox {\boldmath $r$}}
\newcommand{\bfzeta}{\mbox {\boldmath $\zeta$}}
\newcommand{\bfsigma}{\mbox {\boldmath $\sigma$}}

\newcommand{\bfdel}{\mbox {\boldmath $\Delta$}}
\newcommand{\dsdmok}{\frac {d^2\sigma}{dm\,d\Omega_{\bfk}}}
\newcommand{\dsdmoko}{\frac {d^2\sigma_0}{dm\,d\Omega_{\bfk}}}
\newcommand{\dsdmokp}{\frac {d^2\sigma_+}{dm\,d\Omega_{\bfk}}}
\newcommand{\dsdo}{\frac {d\sigma}{d\Omega}}

\newcommand{\aofo}{$a^0_0$$\,\leftrightarrow\,$$f_0$}
\newcommand{\mix}{$a^0_0\,$$-$$f_0$}

\newcommand{\bfmix}{\mbox {\boldmath $a_0\,$$-$$f_0$}}
\newcommand{\pndao}{$pn\to d a^0_0$}
\newcommand{\vpndao}{$\vec pn\to d a^0_0$}

\newcommand{\bfvpp}{\mbox {\boldmath $\vec p p\to d\pi^+\eta$}}
\newcommand{\bfvpn}{\mbox {\boldmath $\vec p n\to d\pi^0\eta$}}
\newcommand{\vpp}{$\vec p p\to d\pi^+\eta$}
\newcommand{\vpn}{$\vec p n\to d\pi^0\eta$}
\newcommand{\ppdkk}{$p p\to d K^+\bar K^0$}
\newcommand{\asym}{A(\theta, \varphi )}

\begin{document}
\preprint{FZJ-IKP-TH-2003-06}
\title{
Angular asymmetries in the reactions \bfvpp\ 
and \bfvpn\ and \bfmix\ mixing }
 
\author{A.E. Kudryavtsev$^1$, V.E. Tarasov$^1$,
J. Haidenbauer$^2$, C. Hanhart$^2$, and J. Speth$^2$}
\address{$^1$Institute of Theoretical and Experimental Physics, \\
B. Cheremushkinskaya 25, 117259 Moscow, Russia \\
$^2$Institut f\"ur Kernphysik, Forschungszentrum J\"ulich,
D-52425 J\"ulich }

\date{\today}
\maketitle

\begin{abstract}
The reactions $pp\to d\pi^+\eta$ and
$pn\to d\pi^0\eta$ are of special interest for 
investigating the $a_0(980)$ ($J^P=0^+$) resonance in  
the process $NN \to da_0 \to d\pi\eta$. We study some 
aspects of those reactions within a general formalism and also 
in a concrete phenomenological model. In particular, 
it is shown that the presence of nonresonant (i.e. without excitation of the 
$a_0$ resonance) contributions to these reactions 
yields nonvanishing values for specific polarization observables,
i.e. to effects like those generated by \mix\ mixing. 
An experimental determination of these observables for the 
reaction $pp\to d\pi^+\eta$ would provide concrete information 
on the magnitude of those nonresonant contributions to $\pi\eta$ production. 
We discuss also the possibility of extracting information 
about \mix\ mixing 
from the reaction \vpn\ with polarized proton beam.
\end{abstract}

\newpage

\section{Introduction}
The reactions \ppdkk\ and $pp\to d\pi^+\eta$ are presently the subject
of experimental investigations by the ANKE collaboration at the COSY accelerator 
in J\"ulich~\cite{1,MB1,VK}. The main issue of this study is to obtain further information
about the scalar $a^+_0$(980) resonance, which decays dominantly into the $K^+\bar K^0$
and $\pi^+\eta$ channels \cite{PDG}. Also a measurement of the production of the 
neutral $a^0_0$ meson in the reaction \pndao\ with the
ANKE spectrometer is planned~\cite{MB2,2,3}. 
The $a^0_0$ production is closely related to the problem of \mix\ mixing \cite{Ach1}. 
A nonzero value for the transition amplitude \aofo\ provides a
forward-backward asymmetry for the reaction \pndao. As was shown in 
ref.~\cite{4}, near the threshold this forward-backward asymmetry is large,
of the order of 10\% -- 15\%. Thus, it is evident that the study of this asymmetry 
can provide useful information on the process of $a_0$$-$$f_0$ mixing.

In a recent paper~\cite{KTHHS} some aspects of the reaction \vpndao\ were
discussed for the case of a perpendicular polarized proton beam. 
Specifically, it was shown that for energies close to threshold 
the angular-asymmetry parameter defined by 
\be
\asym =
\frac{\sigma (\theta, \varphi) -\sigma (\pi-\theta, \varphi+\pi)}
{\sigma (\theta, \varphi) +\sigma (\pi-\theta, \varphi+\pi)}\, ,
\label{1}\ee
with  
\be
\sigma (\theta, \varphi)\equiv \dsdo (\theta, \varphi)\, , 
\label{dXS}
\ee
($\theta$ and $\varphi$ are the polar and azimuthal angles of the
outgoing $\pi\eta$ system in the CMS of the reaction)
is proportional to the \mix\ mixing amplitude. I.e. 
\be
\asym \sim \xi\cdot k\, ,
\label{2}
\ee
where $k$ is the relative momentum of the $a^0_0$ meson with respect 
to the deuteron and $\xi$ is the \mix\ mixing parameter. 
This result is valid in lowest order with respect to the 
momentum $k$, i.e. keeping only contributions that are at most linear 
in $k$. It was shown in ref.~\cite{KTHHS}
that corrections to $\asym$ from isospin-conserving terms 
are of order $k^3$ and, therefore, they are of relevance at higher
energies only. Thus, the study of the angular asymmetry (Eq.~(\ref{1})) 
near the $a^0_0$ threshold
in the reaction \vpndao\ gives information on various invariant
amplitudes for this reaction \cite{KTHHS} but, in particular, on the 
\mix\ mixing parameter $\xi$.

The present paper focusses on the reaction $NN\to d\pi\eta$. 
We take into consideration that the $\pi\eta$ system can not only
be produced via the formation of the $a_0$(980) resonance, i.e.
in an $S$-wave state, but also via other reaction mechanisms
and then can be in a $P$ wave or in even higher partial waves. 
For brevity, in the following we will refer to the latter
contributions as nonresonant or as background of the $NN\to da_0 \to d\pi\eta$ 
amplitude. 
(Note that there can be also a background in the $S$-wave state,
as discussed in Refs.~\cite{Mue,Gri} -- but this is not the issue we 
are concerned with here.) 
Consequences of this nonresonant background contribution for
polarization observables of the reactions \vpp\ and \vpn\ will be discussed.
Specifically, it will be shown that,
because of the presence of the nonresonant background, 
in these reactions the angular-asymmetry
parameter $\asym$ becomes nonzero even without isospin-violation. 
Indeed, like the effect induced by the \mix\ mixing discussed above, 
the contribution of the nonresonant background is also linear in $k$ 
and therefore is difficult to separate from effects of the
\mix\ mixing in the reaction \vpn. 
However, a measurement of this asymmetry in the 
reaction \vpp\ can provide us with information on the contribution of
the nonresonant background to the reaction
$pp\to d a^+_0\to d\pi^+\eta$. 

Recently the reactions \ppdkk\ and $pp\to d\pi^+\eta$ were studied
theoretically in a chiral unitary approach taking into account the
coupling between the 
$K^+\bar K^0$ and $\pi^+\eta$ channels~\cite{Oset}. 
The elementary $\pi\eta$ production amplitude was assumed to be given
by the diagram shown in Fig.~\ref{fig1}a where the $\pi$ and
$\eta$ mesons emerge from different nucleons and rescatter with
each other before being emitted. Since chiral dynamics
suppressed the coupling of the $\pi\eta$ system to $P$-waves
\cite{Meiss}, it follows within this approach that the $\pi^+\eta$ 
system is preferably produced in $S$-waves and that $P$ waves, i.e. the 
nonresonant background discussed above, should be practically 
negligible~\footnote{Note, however, that $P$-waves arise naturally 
in the effective Lagrangian approach of Achasov et al. \cite{Ach2}
based on the anomolous Wess-Zumino action.}. 
Their result motivated us to consider in the present paper 
a different production 
mechanism that does not involve the $\pi\eta$ amplitude directly 
and therefore is not constrained by chiral symmetry. 
We assume that the reaction $pp\to d\pi\eta$ proceeds
via pion exchange between the nucleons followed by the exitation
of the $\Delta$(1232) and the $N^*$(1535) resonances \cite{Gri} which
then produce the $\pi$ and $\eta$ mesons in their respective
decay, cf. Fig.~\ref{fig1}b. Such a reaction mechanism is
certainly suppressed at energies near the $\pi\eta$
threshold where chiral dynamics should be dominating. 
However, in the region of the $a_0$ resonance the
excess energy for the $\pi\eta$ system is already around 
300 MeV. Therefore this mechanism could be already of 
relevance and, specifically, it will introduce $P$-wave contributions. 
We will present corresponding results for the angular symmetry
$\asym$\ and also for differential cross sections.
Clearly, in this context an experimental study of this 
asymmetry parameter is very interesting because it may shed light on the 
validity of the chiral unitary approach~\cite{Oset} 
for the calculation of amplitudes for the reactions \ppdkk\
and $pp\to d\pi^+\eta$ near the $a_0$ threshold. 
  
Finally, we will show that a systematic study
of the angular-asymmetry parameters for both reactions, \vpp\ and \vpn, 
may allow to obtain quantitative information on the \mix\ mixing parameter.

The paper is organized as follows. In Sect. II we provide the
general form of the reaction amplitude for the process $NN \to d\pi\eta$
near threshold where we allow the $\pi\eta$ system to be in an $S$
or a $P$ wave. Furthermore we derive expressions for the corresponding
differential cross sections. In Sect. III we consider a phenomenological
model for the nonresonant ($P$ wave) contributions to the reaction
$pp \to d\pi^+\eta$ and present concrete estimations for the angular symmetry 
$\asym$\ and also for differential cross sections. 
Implications of a possibly non-zero background contribution to the
reaction $NN\to d\pi\eta$ on the issue of \mix\ is discussed in
Sect. IV. The paper ends with a short summary.  

\section{General form of reaction amplitude}

Let us consider first the reaction $pp\to d\pi^+\eta$. If the $\pi^+\eta$
system is produced by the decay of the $a^+_0$ meson it is in an $S$-wave. 
However, if we produce the $\pi^+\eta$ state with an invariant mass of
$m\approx m_{a_0}\approx 980$~MeV but not via the 
$a^+_0$ resonance then the orbital angular momentum of the 
$\pi\eta$ system may be large because, as mentioned, the energy available 
in the $\pi\eta$ system is already around 300 MeV.

We start with the most general form of the amplitude ${\cal M}$ 
for the reaction $NN\to d\pi\eta$. It is given by 
\be
{\cal M} =\phi^T_1\sigma_y\,[F +\bfg\cdot\bfsigma]\,\phi_2 \, ,
\label{3}
\ee
where $\phi^T_1$ and $\phi_2$ are the spinors of the nucleons
($T$ indicates the transposed state vector).
The amplitude in Eq.~(\ref{3}) involves two terms, $F$  and
$\bfg\cdot\bfsigma$, corresponding to the initial total spin
of the nucleons of $S_{NN}=0$ and $S_{NN}=1\,$, respectively.

Note, that both functions $F$ and $\bfg$ are to be linear functions
of the polarization vector $\bfeps^*$ of the outgoing deuteron. 
In the following 
we limit ourselves to the consideration of $S$- and $P$-wave
states for the produced $\pi\eta$ system and, accordingly,
we write both functions $F$ and $\bfg$ as $F=F_S + F_P$ and
$\bfg =\bfg_S +\bfg_P\,$. Taking into account the Pauli principle together
with parity and angular momentum conservation 
and considering only terms that are at most linear in $k$ 
we get 
\begin{eqnarray}
\nonumber
F_S & = & 0\, , \\
\bfg_S & = & a_S\,\bfp\, (\bfk\cdot\bfeps^*)\, +\,
b_S\,\bfeps^*\,(\bfk\cdot\bfp)\, +\, c_S\,\bfk\, (\bfp\cdot\bfeps^*)\,
+\, d_S\,\bfp\,(\bfp\cdot\bfeps^*)\,(\bfk\cdot\bfp)\,
\label{4}
\end{eqnarray}
for the case of the $\pi\eta$ system being in an $S$-wave
(e.g. production via the $a_0$ resonance) and 
\be
F_P = a_0\,(\bfeps^*\cdot [\bfk\times\bfq])\, +\,
b_0\,(\bfeps^*\cdot[\bfp\times\bfq])\,(\bfp\cdot\bfk )\, +\,
c_0\,(\bfeps^*\cdot[\bfp\times\bfk])\,(\bfp\cdot\bfq )\, ,
\label{5}
\ee
\be
\bfg_P  =  a_1\,\bfeps^*\,(\bfp\cdot\bfq )\, +\,
b_1\,\bfp\,(\bfeps^*\cdot\bfq )\, +\, c_1\,\bfq\,(\bfeps^*\cdot\bfp)\,
+\, d_1\,\bfp\,(\bfeps^*\cdot\bfp)\,(\bfp\cdot\bfq)
\label{6}
\ee
for the case of the $\pi\eta$ system being produced in a $P$-wave. 
Here $a_S$, $a_0$, $a_1$, $b_S$, $b_0$, $b_1$, $c_S$, $c_0$, $c_1$,
$d_S$, and $d_1\,$ are independent scalar amplitudes, which may be
considered as being basically 
constants for the near-threshold production ($k$ is small) of the
$\pi\eta$ system with an invariant mass $m=m_{a_0}$ of the $a_0$ meson.
In Eqs.~(\ref{4})-(\ref{6}) we used the following notations:

$\bfp$ is the initial relative momentum in the center-of-mass system
(CMS) of the reaction;

$\bfk$ is the final relative momentum of the deuteron with respect to 
the $\pi\eta$ system in the CMS of the reaction;

$\bfq$ is the relative momentum between the pion and the $\eta$ in the 
$\pi\eta$ CM frame.
\vspace{2mm}

The matrix element ${\cal M}$ (Eq.~(\ref{3})) squared and summed over the
polarizations of the initial neutron is given by 
\be
\overline {|{\cal M}|^2} =\frac{1}{2} \left[ |F|^2\, +\, 2\,{\rm Re}
\left( F^* \bfg\cdot\bfzeta\right)\, +\, (\bfg^* \cdot \bfg)\,
+\, i \left(\bfzeta\cdot [\bfg \times \bfg^{\,*} ]\right)\right]\, ,
\label{7}
\ee
where $\bfzeta$ is the polarization vector of the initial proton, i.e. 
$\bfzeta=\phi^+_1\bfsigma\phi_1\,$. In what follows we shall consider the
vector $\bfp$ to be aligned in the direction of the $z$-axis and the vector 
$\bfzeta$ to be aligned in $x$-direction, so that $\bfzeta\perp\bfp$.

Since we are interested in 
the behaviour of the differential cross section
$d^2\sigma/dm d\Omega_{\bfk}$ and in the angular-asymmetry parameter
$A(\theta, \varphi)$, we are to integrate the expression~(\ref{7})
over the direction of the momentum $\bfq\,$:
\be
\dsdmok = N \int \frac{d\Omega_{\bfq}}{4\pi}\, \overline {|{\cal M}|^2}\, :=
\sigma (m; \theta, \varphi )\, .
\label{8}
\ee
Here $m$ is the invariant mass of the $\pi\eta$ system, $N=kq/(4\pi)^4ps\,$
and $s$ is the square of the total energy in the CMS of the reaction.
After the integration over $d\Omega_{\bfq}$ the $S-P$ interference term
in $\sigma (m; \theta, \varphi)$ disappears. The contribution 
of the $S$-wave part of ${\cal M}$~(\ref{3}) to $\sigma (m; \theta, \varphi)$ 
can be obtained from Eqs.~(22) and (23) of ref.~\cite{KTHHS} and reads
\begin{eqnarray}
\nonumber
\left(\dsdmok\right)_S & = & N\,\biggl\{\,p^2 k^2\, \biggl[
 \frac{1}{2}\,\left(\, |a_S|^2 + |b_S|^2 \,\right)\, \\
\nonumber
 & + & \left[\, |b_S|^2 +\frac{1}{2}\,|b_S + p^2 d_S|^2 +
 {\rm Re}\,\left( a^*_S c_S + (a_S + c_S)^* (b_S + p^2 \,d_S)\right)
 \right]\, \cos^2\theta\,\biggr] \\
 & + & \zeta\,pk\, {\rm Im} (a^*_S b_S + a^*_S c_S + b^*_S c_S
       + p^2 d^*_S c_S )\,\sin\theta\,\cos\theta\,\sin\varphi \,
\biggr\} .
\label{81}
\end{eqnarray}
The contribution from the $P$-wave part
can be deduced from Eqs.~(\ref{5})--(\ref{8}) and amounts to 
\begin{eqnarray}
\nonumber
\left(\dsdmok\right)_P & = & N\,\biggl\{\,\frac{k^2 q^2}{6}\,\Bigl[\,
 2|a_0|^2 + 2|b_0|^2 p^4 \cos^2\theta + |c_0|^2 p^4 \sin^2\theta +
 4{\rm Re}\,(a^*_0 b_0)\, p^2 \cos^2\theta  \\
\nonumber
 & - & 2{\rm Re}\,(a^*_0 c_0)\,p^2 \sin^2\theta \, \Bigr]\, +\,
 \frac{q^2}{3}\,{\rm Re}\,(a^*_0 a_1 - a^*_0 c_1 - c^*_0 a_1 p^2\, )
 \, (\bfzeta\cdot [\bfk\times\bfp ]) \\
\nonumber
 & + & \frac{p^2 q^2}{6}\,\Bigl[\, (3|a_1|^2 + |b_1|^2 + |c_1|^2 )
 +  p^4 |d_1|^2 \\
 & + & 2{\rm Re}\Bigl(\,(a^*_1\, (b_1 + c_1 + p^2 d_1) +
 b^*_1 c_1 + (b^*_1 + c^*_1)\, p^2 d_1 \Bigr)\,\Bigr]\, \biggr\} \, .
\label{9}
\end{eqnarray}
It can be easily seen from Eq.~(\ref{9}) that the only contribution to the 
angular-asymmetry parameter $\asym$, defined in Eq.~(\ref{1}), comes from the 
term proportional to
$\bfzeta\cdot [\bfk\times\bfp ]=\zeta p k\sin\theta \sin\varphi$,
where $\theta$ and $\varphi$ are the polar and azimuthal angles of the
outgoing $\pi\eta$ system in the CMS of the reaction. Hence, because of 
the $P$-wave (nonresonant background) contribution, we get a nonvanishing angular
asymmetry and this asymmetry is linear in $k$. Note that 
$\asym=0$ in the case where the $\pi\eta$ system is produced in an $S$-wave,
as follows from Eq.~(\ref{81}).

Therefore, it is clear 
that the angular asymmetry $\asym$ in the reaction
\vpp\ can provide information on the magnitude of contributions from
partial waves with $l\ge 1$ in the $\pi\eta$ system. 

\section{Estimation of the angular asymmetry for nonresonant
          $\pi^+\eta$ production}

Let us now come to a concrete estimation for the angular-asymmetry
parameter. For that purpose we 
consider a simple model for the reaction $pp\to d\pi^+\eta$ in 
which the $\pi^+\eta$
system can be produced with nonzero internal angular momentum
($l\ge 1$). Specifically, we adopt the one-pion-exchange diagram
in Fig.~\ref{fig1}b. 
We assume that the dominant contribution arises from the 
intermediate $\Delta$(1232) state in the subprocess $\pi N\to \pi N$ 
and from the $S$-wave amplitude of the subprocess $\pi N\to \eta N$. 
The latter is basically given by the contribution of the $N^*$(1535) 
resonance state.
This diagram with the same subprocesses was already used in
ref.~\cite{Gri} in the context of 
the $a_0$ production process in $pN\to d a_0\to d(\pi\eta)$. 
However, in that work the authors were primarily interested in the
$S$-wave part of the one-pion-exchange $\pi\eta$ production amplitude. 
Therefore, the 
angular distribution of the outgoing pion was not calculated in~\cite{Gri}.

The evaluation of this diagram, treating the 
intermediate nucleons and the final deuteron nonrelativistically,
leads to the following expression
for the corresponding $\pi\eta$ production amplitude ${\cal M}$:
\be
{\cal M} = {\cal M}_1+{\cal M}_2\, ,
\phantom{xxx}
{\cal M}_{1,2} =\pm\, A_{1,2}\,\phi^T_{1,2}\sigma_y (\bfeps^*\cdot\bfsigma)\,
(2\bfn_{1,2}\cdot\bfn_2 +
 i\bfsigma\cdot [\bfn_{1,2}\times\bfn_2 ] ) \phi_{2,1}\, ,
\label{101}
\ee
Note that 
the amplitude ${\cal M}$ is antisymmetric with respect to the initial nucleons,
i.e. ${\cal M}_1\leftrightarrow -{\cal M}_2$ when the nucleons are interchanged.
The quantities $A_{1,2}$ are given by 
\be
A_{1,2} = T  \frac{E+m_N}{2m_N\sqrt{2m_N}}\,
 M_{\pi^0 N\to\eta N} M_{\pi N\to\pi N}
\int \frac{d\bfq}{(2\pi)^3} u(q)\, G_{\pi}(t_{1,2})\, .
\label{102}
\ee
Further, 
$2\bfn'_{1,2}\cdot\bfn_2 +
 i\bfsigma\cdot [\bfn'_{1,2}\times\bfn_2 ]\,$ is the spin operator of
the $\Delta$ state in the $\pi N$ scattering amplitude and 
$\bfn'_{1,2}=\bfr_{1,2}/r_{1,2}$ and $\bfn_2 = \bfk_2 /k_2$
are unit vectors. The corresponding momenta, 
$\bfr_{1,2}\!=\!\bfk_2\!+\!\bfq_2\!-\!\bfp_{2,1}$ and
$\bfk_2$, are the 3-momenta of the intermediate (virtual) and final pion 
in the CMS of the reaction, in the notation used
in Fig.~\ref{fig1}b. $T=4\sqrt{2}/3$ is an isospin factor and 
$m_N$ and $E$ are the mass and total CMS energy of the initial proton.
The $S$-wave amplitude for $\pi^0 N\to\eta N$ is obtained from the
relation (see also ref.~\cite{Gri1})
\be
|M_{\pi^0 N\to\eta N}(s_1)|^2=8\pi s_1
\frac{p^{\pi}_{cm}}{p^{\eta}_{cm}}\, \sigma_{\pi^- p\to\eta N}\,
\phantom{xxx}
( s_1=m^2_{\eta N})\, ,
\label{1021}
\ee
where the cross section 
$\sigma_{\pi^- p\to\eta N}\!=\!(21.2\!\pm\! 1.8)p^{\eta}_{cm}$ [$\mu b$]
($p^{\eta}_{cm}$ in MeV$/c$) is taken from the experiment~\cite{Binnie}.
The scalar amplitude for $\pi N\to\pi N$ resulting from the
$\Delta$ resonance reads
\be
M_{\pi N\to\pi N}(m_{\pi N}) =\frac{g^2_{\pi N\Delta}\,
 k_{cm}(\pi')\, k_{cm}(\pi)}
{m_{\pi N } - M_{\Delta} + i\Gamma_{\Delta}/2}\, ,
\phantom{xxx}
g^2_{\pi N\Delta}=
\frac{4\pi M_{\Delta}\Gamma_{\Delta\!\to\!\pi N}}{k^3_R}\, , 
\label{1022}
\ee
where $k_{cm}(\pi')$, $k_{cm}(\pi)$ are the relative momenta
at the $\Delta N\pi$ vertices with the virtual and final pion,
respectively. $k_R$ is the relative momentum in the decay
$\Delta\to N\pi$ at the nominal mass $m_{\pi N}=M_{\Delta}$.

In the following 
all the values which are taken out of the loop
integral~(\ref{102}) are calculated at fixed values
$\bfq_1=\bfq_2=\bfp_d/2$ of the intermediate on-mass-shell nucleons,
where $\bfp_d$ is the momentum of final deutron in the reaction CM.
The loop integral in Eq.~(\ref{102}) contains only the wave function
$u(q)$ of the deuteron and the pion propagator $G_{\pi}(t)$.
We use the $S$-wave part of the deuteron wave function of the full Bonn 
potential~\cite{WFD}. The propagator of the virtual pion, including a
form factor $F_{\pi}(t)$ of monopole type for each $\Delta N\pi$ vertex, 
reads
\be
G_{\pi}(t) =\frac{F^2_{\pi}(t)}{t-\mu^2\!+\!i0},
\phantom{xxx}
F_{\pi}(t)= \frac{\mu^2-\Lambda^2}{t-\Lambda^2},
\label{103}
\ee
where $\mu$ is the pion mass. For the cutoff parameter $\Lambda$ we 
consider the values $\Lambda = 1\div 1.3$~GeV~\cite{Gri}. The four-momentum
transfer squared $t$ for nonrelativistic intermediate
nucleons is given by the relations ($t=t_{1,2}$ for the
corresponding part ${\cal M}_{1,2}$ of the total antisymmetric 
amplitude ${\cal M}$ according to Eq.~(\ref{101}))
\be
t_{1,2}\!-\!\mu^2\!+i0 =
-x\,\left[(\bfq\!-\!\bfdel_{1,2})^2\!-\! a^2_{1,2}\!-\!i0\right]\, ,
\phantom{xxx}
x=(E-\omega_2)/m\, ,
\label{1031}\ee
$$
a^2_{1,2}=\frac{1}{x}\left[\,(T_N-\omega_2)^2 +(\bfp_{2,1}-\bfk_2)^2
\left(\frac{1}{x} -1\right) -\mu^2\,\right]\, ,
\phantom{xx}
\bfdel_{1,2}=\frac{\bfp_{2,1}-\bfk_2}{x} -\frac{\bfp_d}{2}\, .
$$
where $T_N=E-m_N$ and $\omega_2$ is the total energy of the final pion
in the CMS of the reaction.

Let us rewrite the total amplitude ${\cal M}$ (Eq.~(\ref{101})) in the
form of Eq.~(\ref{3}). Then we get
\be
{\cal M} = {\cal M}_1\!+\!{\cal M}_2=
\phi^T_1\sigma_y\,[F\!+\!\bfg\cdot\bfsigma]\,\phi_2\, ,
\phantom{xx}
{\cal M}_{1,2}=\pm\,A_{1,2}\, \phi^T_{1,2}\sigma_y\,[F_{1,2}\!
+\!\bfg_{1,2}\cdot\bfsigma]\,\phi_{2,1}\, ,
\label{104}
\ee
where
$$
F =A_1 F_1 + A_2 F_2\, ,
\phantom{xxx}
\bfg =A_1 \bfg_1 - A_2 \bfg_2\, ,
$$
\be
F_{1,2} =i(\bfeps\cdot[\bfn'_{1,2}\times\bfn_2])\, ,
\phantom{xx}
\bfg_1=2(\bfn'_{1,2}\cdot\bfn_2)\bfeps +
 (\bfn_{1,2}\cdot\bfeps)\bfn_2 -(\bfn_2\cdot\bfeps)\bfn'_{1,2}\, .
\label{105}
\ee

Using Eqs.~(\ref{7}), (\ref{8}) and (\ref{101})-(\ref{105}), we
evaluated the angular asymmetry $A(\theta,\varphi)$, 
defined in Eq.~(\ref{1}), at the angle 
$\theta=90^0$, where the term $2{\rm Re} ( F^* \bfg\cdot\bfzeta )\sim
\bfzeta\cdot [\bfk\times\bfp ]=\zeta p k\sin\theta \sin\varphi$
should produce the maximal effect. The calculations were performed at
$m=980$~MeV (i.e. in the $a_0$-meson region). It turned out that
the resulting angular asymmetry parameter $\asym$ is very small, 
i.e. $A(90^0, \varphi)\le$~1\%. 
In order to understand this we need to go back to the equations
above. Indeed one can immediately see, that 
in the case of zero relative phase
between the quantities $A_1$ and $A_2$ given by Eq.~(\ref{102}), one
should obtain exactly $A(\theta, \varphi) = 0$. This is due to the
relative phase of $90^0$ between the functions $F$ and $\bfg$ in the term
$2{\rm Re} ( F^* \bfg\cdot\bfzeta )$ of Eq.~(\ref{7}) because of the factor
$i$ in $F$ (see Eq.~(\ref{105}) for $F_1$). The latter results from the
spin structure of the $\pi N$ scattering amplitude, which in our case
is given by the excitation of the $\Delta$ resonance alone,
and is connected with the Hermitian form of the interaction
(cf. the factor $i$ in front of the ${\bf \sigma}$ matrix in Eq.~(\ref{101})). 
In fact, the loop integral in Eq.~(\ref{102}) generates an imaginary part from
the on-mass-shell contribution of the exchanged pion. Therefore, due to
permutation of the initial protons, the quantities $A_1$ and $A_2$ acquire
some nonzero relative phase. But still there is only a small effect on the
angular asymmetry.

In this context let us mention that the interaction in the initial
$NN$ system, which is neglected in the simple model calculation presented
here, should also introduce a relative phase between those two amplitudes
and therefore might lead to an enhancement in the predictions for
the angular-asymmetry parameter.

In any case, 
the smallness of this asymmetry effect does not mean that the
$P$-wave fraction in the $\pi\eta$ system is also small.
To illustrate this, we present here some results for the reaction with 
unpolarized proton beam, namely the distribution
$d^2\sigma/dmdz_1$ at selected values of the invariant mass $m$ of the 
$\pi\eta$ system around the $a_0$ mass and for two values of the 
cutoff parameter $\Lambda$ ($\Lambda$ = 1 and 1.3~GeV).
Here $z_1=\cos\theta_1$, where $\theta_1$ is the polar angle of the
outgoing $\pi^+$ meson with respect to the direction of the
proton-beam momentum in the $\pi^+\eta$ rest frame. The calculations
were done at the proton beam energy $T_{lab}=2.65$~GeV.
The resulting angular spectra are shown in the Figs.~\ref{fig2} and 
\ref{fig3}. 
Evidently, 
they are not isotropic but exhibit a strong angular dependence
and, therefore, demonstrate that there are 
significant contributions from higher partial waves ($l\ge 1$).
Thus, a measurement of the angular distribution for the
produced $\pi^+\eta$ system would be rather instructive. In 
particular it should allow to examine the validity of 
the chiral unitary approach used in ref.~\cite{Oset}, which 
implies a completely isotropic angular distribution, 
in the energy region of the $a_0$(980) resonance. 

\section{The reaction \bfvpn\ and the \bfmix\ mixing amplitude}

Let us now consider the reaction $pp\to d\pi^0\eta$. As discussed in
ref.~\cite{KTHHS}, if the $\pi^0\eta$ system is produced in an 
$S$-wave then the only nonzero contribution to the angular-asymmetry 
parameter comes from the \mix\ mixing amplitude. 
Indeed, we have shown there that --
in lowest order in $k$ -- the isospin-violating contribution
to $\asym$ is proportional to $\xi\cdot k$. Thus, an extraction
of the \mix\ mixing parameter $\xi$ would, in principle, 
be feasible from experimental information on the angular asymmetry.

However, as should be clear after the discussion in Sect.~II,
there could be also contributions to $\asym$ from isospin-conserving 
terms, because of the possible presence of $P$ waves in the 
$\pi\eta$ system. 
Such contributions arise from the term in the differential cross 
section proportional to
$\bfzeta\cdot [\bfk\times\bfp ]$. This term is maximal at $\theta$ = 90$^0$ 
and vanishes at $\theta=0^0$. As a consequence, a separation of both 
contributions to $\asym$ is rather complicated.

Still, there is a possibility to separate the contribution
from the isospin-violating part, namely by carrying out a 
combined study of the polarized differential cross sections
for both reactions \vpp\ and \vpn. 
In order to illustrate how this can be achieved, let us first remind 
the reader that the
amplitudes of those reactions are related by 
\be
{\cal M}(pn\to d\pi^0\eta) = \frac{1}{\sqrt{2}}\, 
{\cal M}(pp\to d\pi^+\eta)\, ,
\label{10}
\ee
if isospin is conserved. This relation suggests that one
should consider a ``subtracted'' differential cross section, 
\be
\sigma_{\Delta} (m; \theta, \varphi) :=
\dsdmoko \,-\,\frac{1}{2}\,\dsdmokp \, ,
\label{11}
\ee
where $\sigma_0$ and $\sigma_+$ are the cross sections of the processes
\vpn\ and \vpp, respectively. Evidently, the angular asymmetry
$\asym$, evaluated according to Eq.~(\ref{1}) for this
``substracted'' differential cross section
$\sigma_{\Delta} (m; \theta, \varphi )$ 
does not contain terms induced by isospin-conserving processes and 
hence provides only information on isospin violating effects.
Note that in the simple model considered in Sect.~III the angular
asymmetry in the reaction \vpp\ is small in any case and, therefore, 
one might expect that any effects seen in the experiment  
should come mainly from isospin-violating \mix\ mixing.

\section{Summary}

We presented a discussion on effects of background contributions to 
differential cross sections and the angular asymmetry $\asym$ for the 
reactions $pp\to d\pi^+\eta$ and $pp\to d\pi^+\eta$. Specifically, we pointed 
out that already in lowest order in the relative momentum between the
deuteron and the $\pi\eta$ system
the angular asymmetry $\asym$ can be nonzero 
-- even without isospin mixing effects -- 
because of the presence of higher partial waves in the $\pi\eta$ system. 
As a consequence, a measurement of the angular asymmetry
$\asym$ with polarized proton beams for the reaction \vpp\ should
provide useful information on the role of higher partial waves 
in this reaction. It would also allow to examine the  
validity of model calculations of the reaction 
$pp\to d\pi^+\eta$ based on chiral constraints~\cite{Oset}
for energies around the the $a_0$ threshold. 

We also argued that a
combined analysis of differential cross sections for both reactions,
\vpp\ and \vpn, may facilitate the  extraction of isospin-violating effects
induced by \mix\ mixing and allow to shed light on the nature of these
scalar mesons.

\section*{Acknowledgements}

The authors are thankful to V.V. Baru and M. B\"uscher for useful
discussions.
This work was partly supported by the DFG-RFBI Grant
No.~02-02-04001 (436 RUS 113/652/1-1) and by the RFBR Grant
No.~00-15-96562.


\newpage

\begin{figure}
\caption{Nonresonant mechanisms for the reaction $pp\to d\pi^+\eta$. 
(a) $\pi\eta$ rescattering.
(b) Pion-exchange diagram involving the $\pi N \to \Delta \to \pi N$ and 
$\pi N \to \eta N$ reaction amplitudes.
}
\label{fig1}
\end{figure}

\begin{figure}
\caption{
 Angular distributions of the outgoing $\pi^+$ meson in the $\pi^+\eta$
 rest frame for several values $m$ of the $\pi^+\eta$ invariant mass
 in the $a_0$-mass region. The dashed, solid and dotted curves correspond
 to $m$ = 950, 980 and 1020 MeV$/c^2$, respectively. The calculations are
 performed at $T_{lab}=2.65$~GeV and for a cutoff parameter~(\ref{103})
 $\Lambda=1$~GeV$/c$.}
\label{fig2}
\end{figure}

\begin{figure}
\caption{
The same distributions as in Fig.~\protect\ref{fig2} but 
for $\Lambda=1.3$~GeV$/c$.
}
\label{fig3}
\end{figure}

\newpage
\vglue 1cm
\begin{figure}
\begin{center}
\epsfig{file=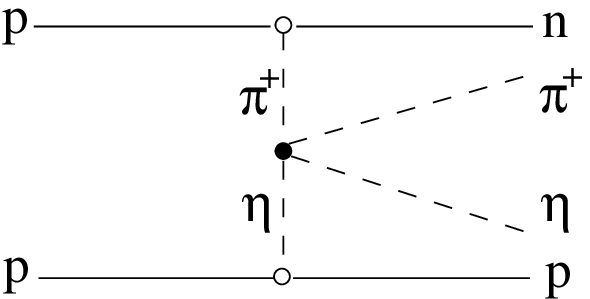,width=7.0cm}
\end{center}
\end{figure}

\vskip 1cm 
\center{Fig. 1a}

\begin{figure}[tb]
\vspace{5cm}
\includegraphics{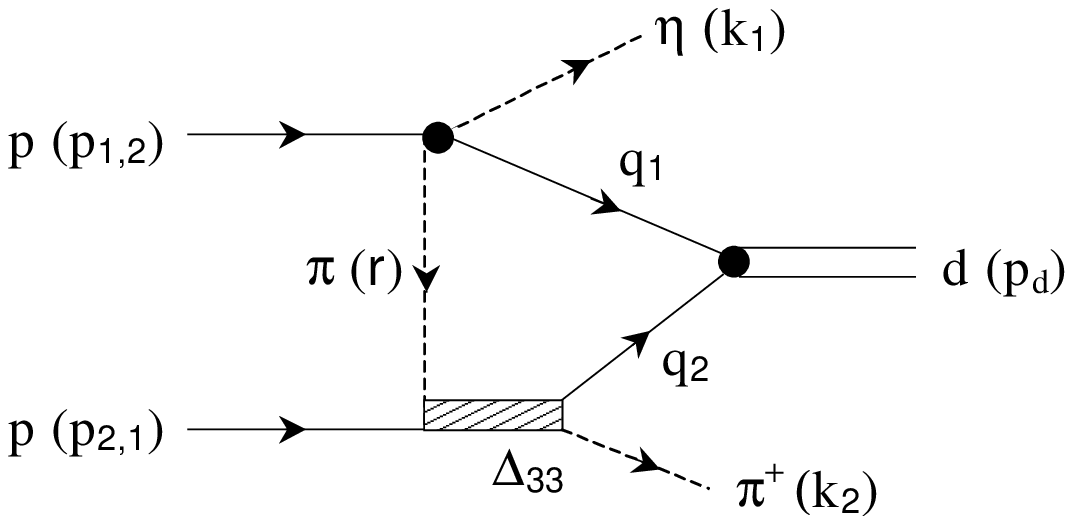}
\end{figure}

\vskip 5cm 
\center{Fig. 1b}

\newpage
\vglue 1cm
\begin{figure}[tb]
\vspace{5cm}
\includegraphics{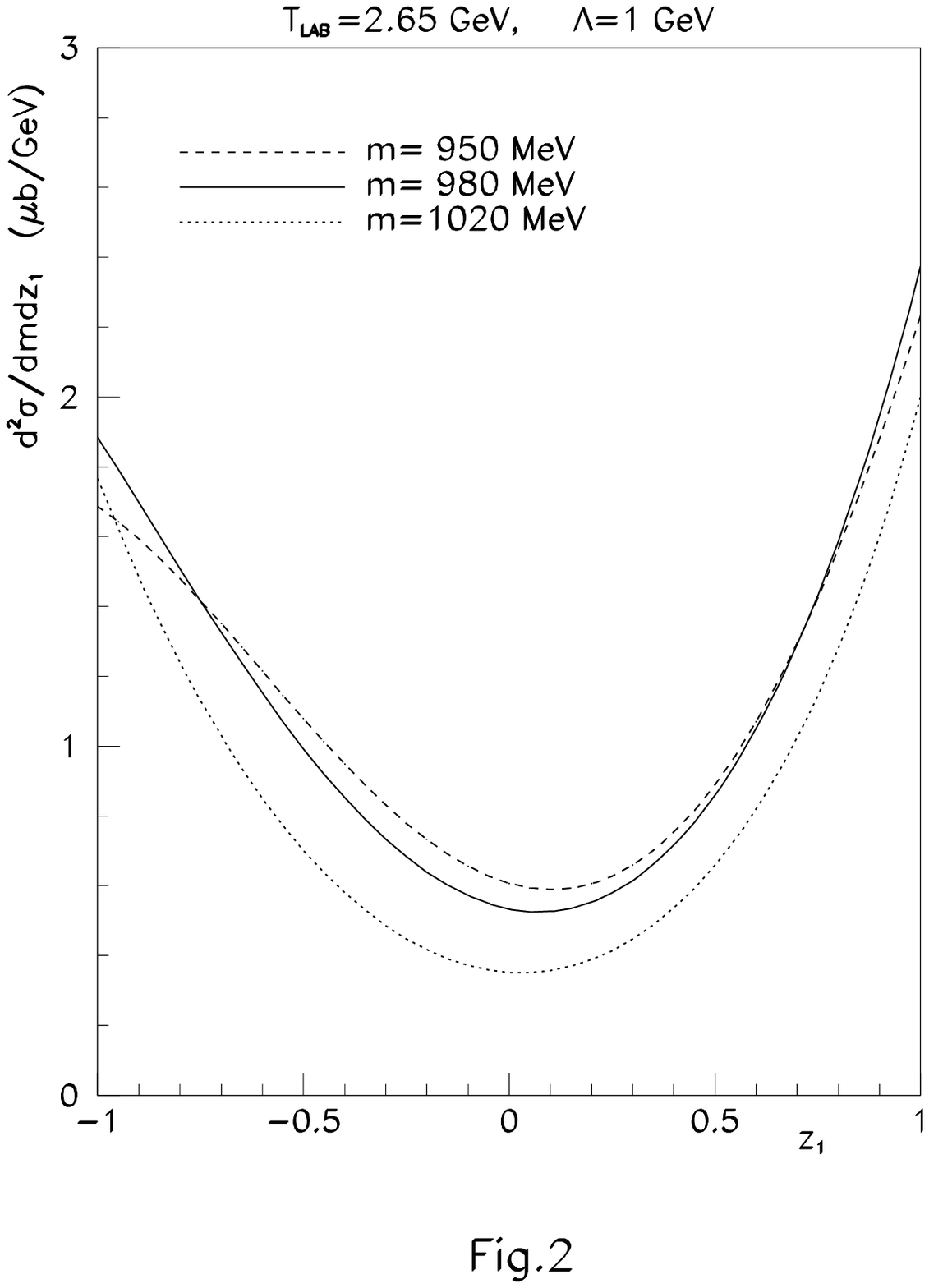}
\end{figure}

\newpage

\vglue 1cm
\begin{figure}[tb]
\vspace{5cm}
\includegraphics{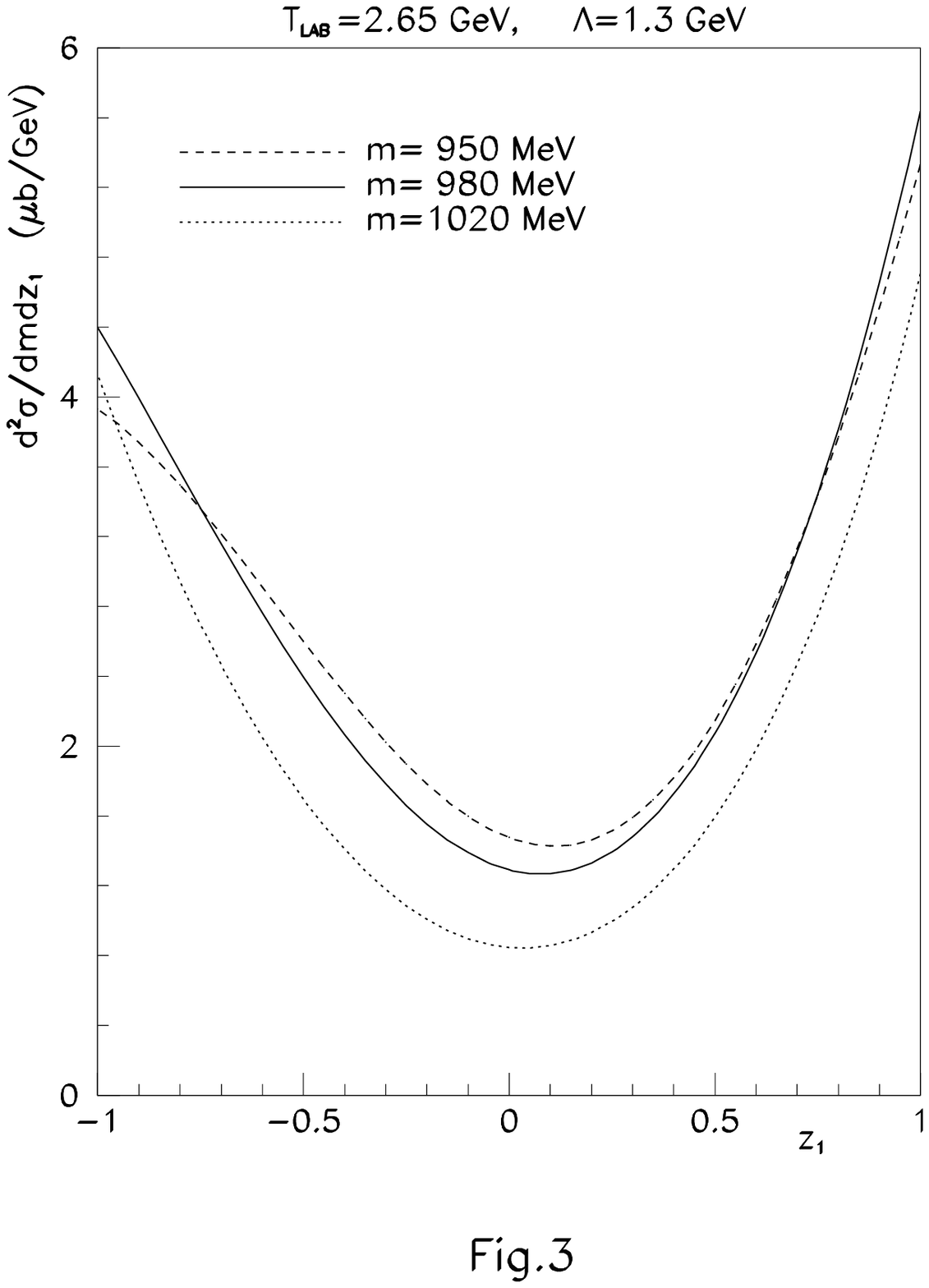}
\end{figure}


\begin{thebibliography}{99}
%
\bibitem{1} Proceedings of the Workshop on $a_0$(980) physics with ANKE, 
            edited by M.~B\"uscher and V.~Kleber, 
            Berichte des
            Forschungszentrums J\"ulich J\"ul-3801 (2000), pp. 3-34.
%
\bibitem{MB1} M. B\"uscher, F.P. Sassen, N.N. Achasov, and L. Kondratyuk,
hep-ph/0301126. 
%
\bibitem{VK} P. Fedorets and V. Kleber, 
Proceedings of the Conference on Quarks and Nuclear Physics (QNP 2002), 
J\"ulich, Germany, 9-14 June 2002, Eur. Phys. J. {\bf A}, in print. 
%
\bibitem{PDG}   C. Caso {\it et al.}, {\it Review of Particle Physics},
                Eur. Phys. J. {\bf C 3}, 1 (1998).
%
\bibitem{MB2} M. B\"uscher et al., 
http://www.fz-juelich.de/ikp/anke/doc/anke/proposal/a0-f0.ps.gz 
%
\bibitem{2} V.E.~Tarasov and A.E.~Kudryavtsev,    
            Proceedings of the Workshop on $a_0$(980) physics with ANKE, 
            edited by M.~B\"uscher and V.~Kleber, 
            Berichte des
            Forschungszentrums J\"ulich J\"ul-3801 (2000), p. 151.
%
\bibitem{3} Proceedings of the 2nd ANKE 
            Workshop on ``Strangeness Production in $pp$ and
            $pA$ Interactions at ANKE'', 
            edited by M.~B\"uscher, V.~Kleber, P.~Kulessa and M.~Nekipelov, 
            Berichte des Forschungszentrums
            J\"ulich J\"ul-3922 (2001), pp. 209-228.
%
\bibitem{Ach1} N.N. Achasov, S.A. Devyanin, and G.N. Shestakov, Phys. Lett.
B {\bf 88}, 367 (1979); Yad. Fiz. {\bf 33}, 1337 (1981)
[Sov. J. Nucl. Phys. {\bf 33}, 715 (1981)]. 

\bibitem{4} A.E.~Kudryavtsev and V.E.~Tarasov, JETP Lett. {\bf 72}, 401
            (2000); nucl-th/0102053.
%
\bibitem{KTHHS} A.E.~Kudryavtsev, V.E.~Tarasov, J.~Haidenbauer,
                C.~Hanhart, and J.~Speth,
                Phys. Rev. C {\bf 66}, 015207 (2002).
%
\bibitem{Mue} H. M\"uller, Eur. Phys. J. A {\bf 11}, 113 (2001). 
%
\bibitem{Gri} L.A.~Kondratyuk, E.L.~Bratkovskaya, and V.Yu.~Grishina,
             {\tt nucl-th/0207033}, Phys. Atom. Nucl. {\bf 66}, 152
             (2003). 
%
\bibitem{Oset} E.~Oset, J.~A.~Oller, and Ulf-G.~Mei{\ss}ner,
               Eur. Phys. J. {\bf A12}, 435 (2001).
%
\bibitem{Meiss} V.~Bernard, N. Kaiser, and Ulf-G.~Mei{\ss}ner,
               Phys. Rev. D {\bf 44}, 3698 (1991).
%
\bibitem{Ach2} N.N. Achasov and G.N. Shestakov, Phys. Rev. D {\bf 63},
014017 (2000); Yad. Fiz. {\bf 65}, 579 (2002) 
[Phys. Atom. Nucl. {\bf 65}, 552 (2002)]. 
\bibitem{Gri1} V.Yu.~Grishina {\it et al.},
               Phys. Lett. B {\bf 475}, 9 (2000).
%
\bibitem{Binnie} D.M.~Binnie {\it et al.},
               Phys. Rev. D {\bf 8}, 2789 (1973).
%
\bibitem{WFD} R.~Machleidt, K.~Holinde, and Ch.~Elster,
              Phys. Rep. {\bf 149}, 1 (1987).
%
%
\end{thebibliography}
\end{document}